\documentclass[a4paper,fleqn]{cas-dc}

\usepackage[numbers,sort&compress]{natbib}
\usepackage{placeins}
\usepackage{caption}
\usepackage{hyperref}
\begin{document}
\let\WriteBookmarks\relax
\def\floatpagepagefraction{1}
\def\textpagefraction{.001}
\let\printorcid\relax 

\shorttitle{}    
\shortauthors{}
\title [mode = title]{Wireless Environmental Information Theory: A New Paradigm towards 6G Online and Proactive Environment Intelligence Communication
}  

%

%

\author[1]{Jianhua Zhang}

\author[1]{Li Yu}
\cormark[1]
\ead{li.yu@bupt.edu.cn}

\author[1]{Shaoyi Liu}

\author[1]{Yichen Cai}

\author[1]{Yuxiang Zhang}

\author[1]{Hongbo Xing}

\author[2]{Tao jiang}
%
%

\affiliation[1]{organization={State Key Laboratory of Networking and Switching Technology},
	addressline={Beijing University of Posts and Telecommunications}, 
	city={Beijing},
	postcode={100876}, 
	country={China}}

\affiliation[2]{organization={Future Research Laboratory},
	addressline={ China Mobile Research Institute}, 
	city={Beijing},
	postcode={100032}, 
	country={China}}
\cortext[1]{Corresponding author}

\renewcommand{\textfraction}{0} 
\begin{abstract}
	The channel is one of the five critical components of a communication system, and its ergodic capacity is based on all realizations of statistic channel model. This statistical paradigm has successfully guided the design of mobile communication systems from 1G to 5G. However, this approach relies on offline channel measurements in specific environments, and the system passively adapts to new environments, resulting in deviation from the optimal performance. With the pursuit of higher capacity and data rate of 6G, especially facing the ubiquitous environments, there is an urgent need for a new paradigm to combat the randomness of channel, i.e., more proactive and online manner. Motivated by this, we propose an environment intelligence communication (EIC) based on wireless environmental information theory (WEIT) for 6G. The proposed EIC architecture is composed of three steps: Firstly, wireless environmental information (WEI) is acquired using sensing techniques. Then, leveraging WEI and channel data, artificial intelligence (AI) techniques are employed to predict channel fading, thereby mitigating channel uncertainty. Thirdly, the communication system autonomously determines the optimal air-interface transmission strategy based on real-time channel predictions, enabling intelligent interaction with the physical environment. To make this attractive paradigm shift from theory to practice, we answer three key problems to establish WEIT for the first time. How should WEI be defined? Can it be quantified? Does it hold the same properties as statistical communication information? Furthermore, EIC aided by WEI (EIC-WEI) is validated across multiple air-interface tasks, including channel state information (CSI) prediction, beam prediction, and radio resource management. Simulation results demonstrate that the proposed EIC-WEI significantly outperforms the statistical paradigm in decreasing overhead and performance optimization. Finally, several open issues are discussed, including its accuracy, complexity, and generalization. This work explores a novel and promising way for integrating communication, sensing, and AI capability in 6G.

\end{abstract}


\maketitle

\section{Introduction}\label{secI}

\begin{figure*}[!t]
	\centerline{\includegraphics[width = 17cm]{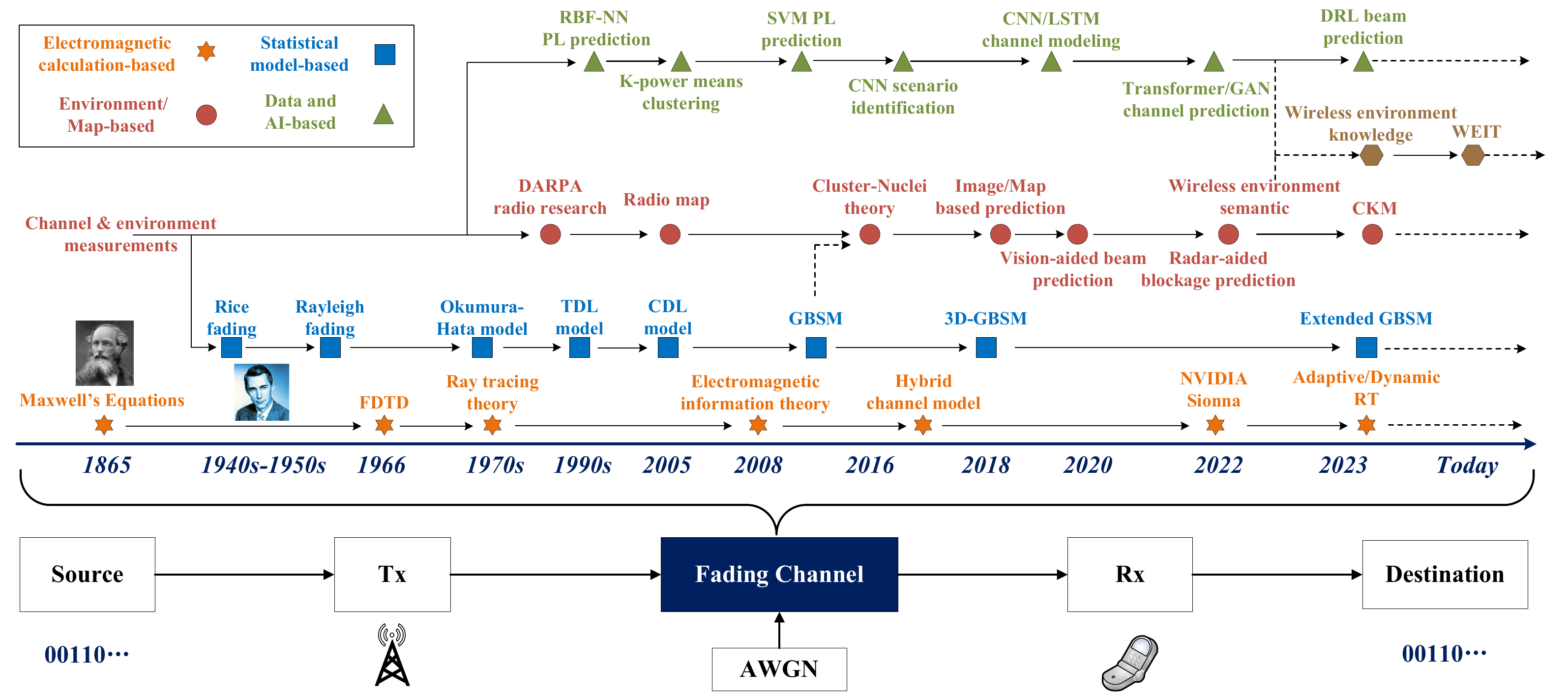}}
	\caption{The timeline of wireless channel research by different methodologies.}
	\label{route}
\end{figure*}

In 1948, C. E. Shannon published his seminal paper, \emph{“A Mathematical Theory of Communication,”} in which he defined communication systems as comprising five components: information source, transmitter, receiver, destination, and channel. He initially describes the channel as: \emph{“the medium used to transmit the signal from transmitter to receiver. It may be a pair of wires, a coaxial cable, a band of radio frequencies, a beam of light, etc.} \cite{Shannon}” It is evident that the aforementioned mediums are all physical elements. In order to consider certain general problems involving communication systems, he elegantly represents the various elements as mathematical entities—random variables and statistical processes, idealized from their physical counterparts.
Mobile communication uses \emph{``a band of radio frequencies"} as the physical carrier, and its channel entity is also stochastically modeled as Shannon’s thought. It has experienced five generations since the 1980s with the revolutionary evolution in services and techniques. Simultaneously, the radio channel as the carrier entity of each generation of mobile communication, its statistical model’s dimension and parameters are also dramatically expanded, as the statistical model-based given in Fig. \ref{route} \cite{engineering1,engineering2,6G1,6G2}. Specifically, the channel for 1G is modeled as a time domain and one-dimensional random process, and the distribution of random variables, including the amplitude and phase of the multipath component (MPC) are key parameters to determine it. In the 2G era, as the channel bandwidth increases, the MPC can be resolved in one delay bin and the classical tap-delay line (TDL) model appears. It is time-frequency two dimensions and the added parameter is delay. Around 1996, MIMO arises to take advantage of the channel’s spatial freedom to increase system capacity, then spatial channel model (SCM) for 3G by 3GPP and geometry-based stochastic model (GBSM) for 4G by ITU are standardized. Those statistical models are time-frequency-horizontal angle of three dimensions, and the added random parameters are angles of arrival/departure at the azimuth plane. Around the 2010s, the active antenna array draws researchers’ attention for flexibly realizing three-dimensional (3D) beamforming, and the theoretical proof of 3D MIMO capacity increases as the vertical MPC dispersion larger \cite{CLcapacity,3Dmodel}. Then angles of arrival/departure at the vertical plane are introduced for the 5G channel model, and 3D-GBSM is a time-frequency-horizontal angle-elevation angle in four dimensions \cite{3D_MIMO1, 3D_MIMO2}.

The above evolution of channel models from 1G to 5G is the success of the statistical paradigm for mobile communication. Previous researchers also realize that a purely statistical model cannot reflect different environmental influences, like the building density, scatter types, height of antennas, etc. In order to overcome this, an experimental method is widely used historically and we name it as the two-step method. Firstly, measurements are implemented with the channel sounder and the amount of data is collected in one typical mobile application environment. Second, the distribution of random parameters, including delay, angle, polarization, etc., are fitted from channel impulse response (CIR) data. Moreover, the parameters to describe those distributions, including the mean and variance, are given one-by-one environment. 
Even two-step experiment-based statistical models are better than the pure theoretical-based models and are mainstream of standardization, like 3GPP TR 38.900/901 and ITU-R M.2412 for 5G \cite{ITU_2412,3Gpp38901}. However, they are still statistical models with a limited type of test environments, such as Urban Macrocell (UMa), Urban Microcell (UMi), Indoor Hotspot (InH), and Rural Macrocell (RMa) \cite{3Gpp38901}. Moreover, we can find that the data used for distribution fitting are offline collected. When those offline statistical models are used to design and test the performance of new techniques, systems have to passively adapt to the various new environments in practice, which often results in their deviation from the optimal performance. Not mention to those statistical models can’t consider the real-time uncertainty of environmental influence, like newly appeared buildings, vehicles, bodies, etc. The system performance will be degraded by those dynamic blockers and the offline model will lead them blind to those variations. In all, the offline statistical model paradigm often requires both time and money costs for network optimization in real environments.

In June 2023, ITU-R defines the primary objectives and six typical application scenarios for 6G. The new scenarios include integrated sensing and communication (ISAC), artificial intelligence (AI) and communication, and ubiquitous connectivity, in addition to the three expanded 5G scenarios \cite{ITU_6G, 6G3, 6G4, 6G5, 6G6, 6G7, 6G8, 6G9}. Moreover, with the further enhancement of data rates and capacity in 6G, traditional communication models face increasing challenges, especially in ubiquitous environments. To combat the randomness of the 6G channel, there is an urgent need for a new paradigm, one that adopts a more proactive and online approach. Fortunately, either base stations or terminals in 6G increasingly incorporate diverse devices, including industrial sensors, depth cameras, millimeter-wave radar, etc. \cite{ISAC70}. The earlier research has already considered the environment through sensing technologies. As environment/map-based methodology in Fig. \ref{route}, in the 1990s, the Defense Advanced Research Projects Agency (DARPA) pioneered research in radio map and the radio signal power is calculated based on the grid map information systems.with multiple sensor-equipped nodes, including vehicles, aircraft, and drones \cite{DARPA}. Moreover, depth camera images and computer vision (CV) techniques are also utilized to reconstruct real-time 3D environments, and then match stochastic clusters with corresponding environment scatterers, forming semi-deterministic cluster-nuclei to improve the accuracy of statistical channel model \cite{cluster_nuclei,cluster_nuclei2,patent1,patent2}. In addition, RGB images gotten by cameras are used to relate positions of scatter with mobility of terminal, to achieve proactive beam selection \cite{vision}. Meanwhile, radar sensors can also be used to obtain information about the distance, velocity, and angle of moving objects, allowing the beam to avoid forward blockers \cite{radar}. In all, different sensing methods have been utilized to collect a kind of environmental information to help the system "see" or "hear" the real-time varying environments. For generality, we give a definition to wireless environmental information (WEI), that is the environment scatters physical description and properties (such as geometric size, mobility, material types, etc.) that can help to eliminate channel uncertainty and affect MPC variation (such as phase, delay, angle, etc.) for the wireless communication system. However, as a kind of information, what properties it has, can we use entropy to quantize it? These urgent questions remain unanswered.

Meanwhile, with the rapid advancement and extensive spread of AI technologies, particularly machine learning (ML), it is expected 6G will be more intelligent \cite{ITU_6G}. Especially, there are some works to introduce AI methods for combating channel uncertainty and randomness \cite{DL1,DL2}. For example, a deep convolutional neural networks (CNN) based approach is designed to predict path loss exponent and shadowing factor, directly from 2D satellite images \cite{satellite_map1}. An environment feature extraction module based on CNN is proposed to process panoramic environment images to predict accurate CSI with lower pilot overhead \cite{slz}. Channel knowledge map (CKM) is proposed as a site-specific database, tagged with the locations and channel-related information to facilitate CSI acquisition \cite{CKM}. The above works have investigated the relationship between environmental information and channels, via the ML methods. However, to further improve system performance, specific transmission tasks should be taken into consideration. Take beam prediction as a task, environment semantic is abstracted to form environment sensing, channel prediction and task application loop \cite{semantic}.

\begin{table*}[!t]
	\setlength{\extrarowheight}{0pt} 
	\renewcommand{\arraystretch}{1.10} 
	\caption{The important milestones of channel research progress from environmental perspective.}
	\begin{tabular}{c|c|c|c|c}
		\hline
		\textbf{Methodology}                                        & \textbf{Year(s)}     & \textbf{\makecell{Environment \\type }}                & \textbf{\makecell{Online/Offline \\ application}} &\textbf{ Representative   progress }                \\ \hline
		\multirow{7}{*}{\makecell{Electromagnetic \\ calculation based}} & 1865       & N/A                       & N/A                       & Maxwell's   equations \cite{Maxwell}                  \\ \cline{2-5} 
		& 1966       & Deterministic & N/A                       & FDTD  method  \cite{FDTD,FDTD2,FDTD3}                           \\ \cline{2-5} 
		& 1960s      & Deterministic & Offline                       & Ray   tracing theory \cite{ray_tracing1, ray_tracing2,ray_tracing3,ray_tracing4}                     \\ \cline{2-5} 
		& 2008       & N/A                       & N/A                       & Electromagnetic   information theory \cite{Mikki,EIT2016,EIT2023,EITtrans,horse}      \\ \cline{2-5} 
		& 2010s      & Deterministic  & Offline                   & Hybrid   channel model\cite{hybird1,hybird2,hybird3}                    \\ \cline{2-5} 
		& 2022      & Deterministic  & Offline                   & NVIDIA Sionna \cite{sionna}                   \\ \cline{2-5} 
		& 2024       & Deterministic & Offline                    & Adaptive/Dynamic   RT \cite{RT1,RT2, RT3}                    \\ \hline
		\multirow{7}{*}{\makecell{Statistical\\ model based}}           & 1944       & Statistical     & Offline                   & Rice   fading  \cite{Rice}                           \\ \cline{2-5} 
		& 1968       & Statistical  & Offline                   & Rayleigh   fading \cite{Rayleigh}                        \\ \cline{2-5} 
		& 1980       & Statistical   & Offline                   & Okumura-Hata   model  \cite{Hata,Hata2}                    \\ \cline{2-5} 
		&       1997     & Statistical  &           Offline       & TDL model  \cite{TDL1, TDL2}                                    \\ \cline{2-5} 
		& 2005       & Statistical & Offline                   &  Cluster delay line (CDL) model \cite{3Gpp25996}             \\ \cline{2-5} 
		& 2010s      & Statistical    & Offline                   &  Geometry-based stochastic model  \cite{3Gpp36873,ITU_2135,3Gpp38901}                 \\ \cline{2-5} 
		& 2020s       & Statistical     & Offline                    & Extended GBSM    \cite{6Gmodel1,6Gmodel2}                      \\ \hline
		\multirow{7}{*}{\makecell{Environment/map \\  based}}         & 1990s       & Deterministic  & Offline                  & DARPA radio research    \cite{DARPA}                       \\ \cline{2-5} 
		& 2000s      & Deterministic  & Offline                  & Radio map  \cite{REM,REM1}                   \\ \cline{2-5} 
		& 2016       & Deterministic  & Offline                    & Cluster-nuclei  theory \cite{cluster_nuclei,cluster_nuclei2}                  \\ \cline{2-5} 
		& 2017       & Deterministic  & Offline                    & Image/Map based prediction \cite{satellite_map1,satellite_map3}                  \\ \cline{2-5} 
		& 2020s       & Deterministic   & Offline                    & Vision/Radar-aided prediction   \cite{vision,radar}                     \\ \cline{2-5} 
		& 2022       & Deterministic  & Online                       & Wireless environment semantic \cite{semantic}     \\ \cline{2-5} 
		& 2023       & Deterministic   & Online                    & Channel knowledge map   \cite{CKM,CKM2,CKM3,CKM4}                \\ \hline
		\multirow{7}{*}{Data and AI based}                 & 1990s       & Deterministic   & Offline                   & RBF-NN path loss prediction   \cite{RBF}                  \\ \cline{2-5} 
		&     2005      & Deterministic   & Offline                   & K-power means clustering \cite{clustering}             \\ \cline{2-5} 
		&     2013       & Deterministic   & Offline                   &  SVM path loss prediction \cite{SVM}             \\ \cline{2-5} 
		&     2017       & Deterministic   & Online                   &  CNN scenario identification \cite{CNNscenario}             \\ \cline{2-5} 
		& 2018       & Deterministic    & Online                   & CNN/LSTM channel modeling \cite{LSTM}         \\ \cline{2-5} 
		& 2022       & Deterministic  & Online                    &  Transformer/GAN channel prediction  \cite{GAN,Trans}                                    \\ \cline{2-5} 
		& 2023 & Deterministic   & Online                    & DRL beam prediction \cite{DRL} \\ \hline
		\multirow{2}{*}{\makecell{Environment \\ intelligence based}}                 & 2023       & Generalized   & Online                   & Wireless environment knowledge pool  \cite{REKP, cyc, FITEE}                  \\ \cline{2-5}
		& This paper & Generalized   & Online                    & Wireless   environmental information theory \\ \hline
	\end{tabular}
\label{routetable}
\end{table*}


The above progress of integrating communication, sensing, and AI capability in 6G, especially recently Feifei Li's spatial intelligence to stress the interaction between 3D space and human/machine inspired us \cite{lifeifei}. This paper first propose a new paradigm for 6G, named as environment intelligence communication (EIC). EIC involves three key steps: First, leveraging sensing techniques to acquire accurate and extensive environmental information; second, predicting channel variations based on WEI; and third, enabling the communication system to autonomously determine the optimal transmission strategy, achieving intelligent interaction. Through EIC, 6G system will be online and proactive to react environment uncertainty. 
To support EIC, wireless environmental information theory (WEIT) is proposed, introducing the definition, classification, and properties of WEI, along with its relationship to environmental entropy for the first time. The feasibility of WEI-assisted channel prediction is investigated, the EIC-WEI architecture is introduced, and validate its performance gains across various tasks. Finally, several open issues are discussed, including its accuracy, complexity, and generalization.

\section{Wireless environmental information theory}\label{secII}

As mentioned earlier, the propagation environment and the wireless channel are inextricably linked. We will begin by reviewing research on the relationship between the environment and the channel, tracing the evolution of understanding regarding the significance of the environment. This leads us to introduce our key concept—WEI, and subsequently discuss the relationship between channel uncertainty and wireless environmental entropy, forming the wireless environmental information theory.

\subsection{Propagation environment and channel related work}

As illustrated in Fig. \ref{route}, channel research can be classified using different methodologies. Based on the interaction between the environment and the channel, it can be divided into four categories: the first is electromagnetic calculation-based, the second is statistical model-based, the third is environment/map-based, and the fourth is data and AI-based. 

\textbf{Electromagnetic calculation-based:} The electromagnetic calculation originated from J. C. Maxwell theoretically provides a perfect mathematical description of electromagnetic waves in 1865, by constraining the boundary conditions of electric and magnetic fields, the propagation of electromagnetic waves can be rigorously solved \cite{Maxwell}. However, analytical solutions are limited to simple geometries with well-defined boundaries. The finite-difference time-domain (FDTD) method simplifies solving Maxwell’s equations by discretization, enabling numerical solutions \cite{FDTD,FDTD2,FDTD3}. Ray tracing (RT) rooted in geometric optics, models electromagnetic wave propagation in real-world environment by computing multipath effects such as direct paths, reflections, and scattering \cite{ray_tracing1,ray_tracing2,ray_tracing3,ray_tracing4}. Electromagnetic information theory (EMIT) extends computational electromagnetics by integrating circuit and antenna effects to enhance communication degrees of freedom \cite{horse,Mikki,EIT2016,EIT2023,EITtrans}. Recently, by integrating prior environmental information, RT improves accuracy and reduces complexity, establishing itself as a widely used deterministic channel modeling method \cite{RT1,RT2,RT3}.

\textbf{Statistical model-based:} Despite their high accuracy, high computational complexity consistently challenges electromagnetic calculation based channel research methods. Statistical models, with their strong adaptability and lower computational demands, have become the mainstream in standards. As summarized in the introduction, standardized channel models have evolved from one-dimensional models, such as the Hata model \cite{Hata,Hata2,Rayleigh,Rice}, to two-dimensional broadband channel models, and eventually to three-dimensional multi-parameter models like the 3D-GBSM \cite{3Gpp25996,3Gpp36873,3Gpp38901}. In 6G systems, to flexibly support new technologies such as ISAC, extremely large-scale MIMO (XL-MIMO), and reconfigurable intelligent surface (RIS), while maintaining backward compatibility with 5G models, the extended GBSM facilitates this smooth transition \cite{6Gmodel1,6Gmodel2}.

\textbf{Environment/Map-based:} Channel studies based on environment and map have become a feasible solution, enabled by the advancements in depth camera,  3D point cloud light detection and ranging (LiDAR), global positioning system (GPS) and other sensing  technologies. Leveraging extensive measurement data, the concept of the radio map (RM) has been introduced to address the limitations of traditional static spectrum resource allocation effectively \cite{REM,REM1}. Satellite imagery and map data can predict channel parameters such as shadowing factors \cite{satellite_map1, satellite_map3}. Additionally, CKM is an evolving concept that integrates sensing capabilities to create detailed spatiotemporal maps containing channel characteristics, supporting channel behavior prediction and communication performance optimization \cite{CKM, CKM2, CKM3, CKM4}.

\textbf{Data and AI-based:} AI demonstrates exceptional capability in uncovering hidden patterns from extensive channel data, driving significant advancements in channel research based on data and AI. Certain studies employ CNNs for scenario identification and assist in channel modeling \cite{CNNscenario,LSTM}. Using transformer networks, a joint time-frequency-space domain channel prediction method is proposed, through which the cross-domain strategy can accelerate the training process of the AI model and improve the prediction accuracy \cite{Trans}. To address dynamic beam selection, a deep reinforcement learning based method is presented to cope with the change of wireless channel environment \cite{DRL}. By integrating the reasoning capabilities of AI and wireless environments, wireless environment knowledge (WEK) establishes mapping relationships between wireless environments and channel characteristics \cite{REKP, FITEE,cyc}.


Table \ref{routetable} presents a summary of the timeline, connections, and application differences of various research methodologies from environmental perspectives. It is evident that research on channels is gradually shifting towards environment and intelligence. Environment intelligence communication represents the next step in the transition of channels from offline to online approaches. To better support EIC, the aforementioned research motivates the proposal of wireless environmental information theory, which is the focus of this section.

\begin{figure*}[!t]
	\centerline{\includegraphics[width = 5.5in]{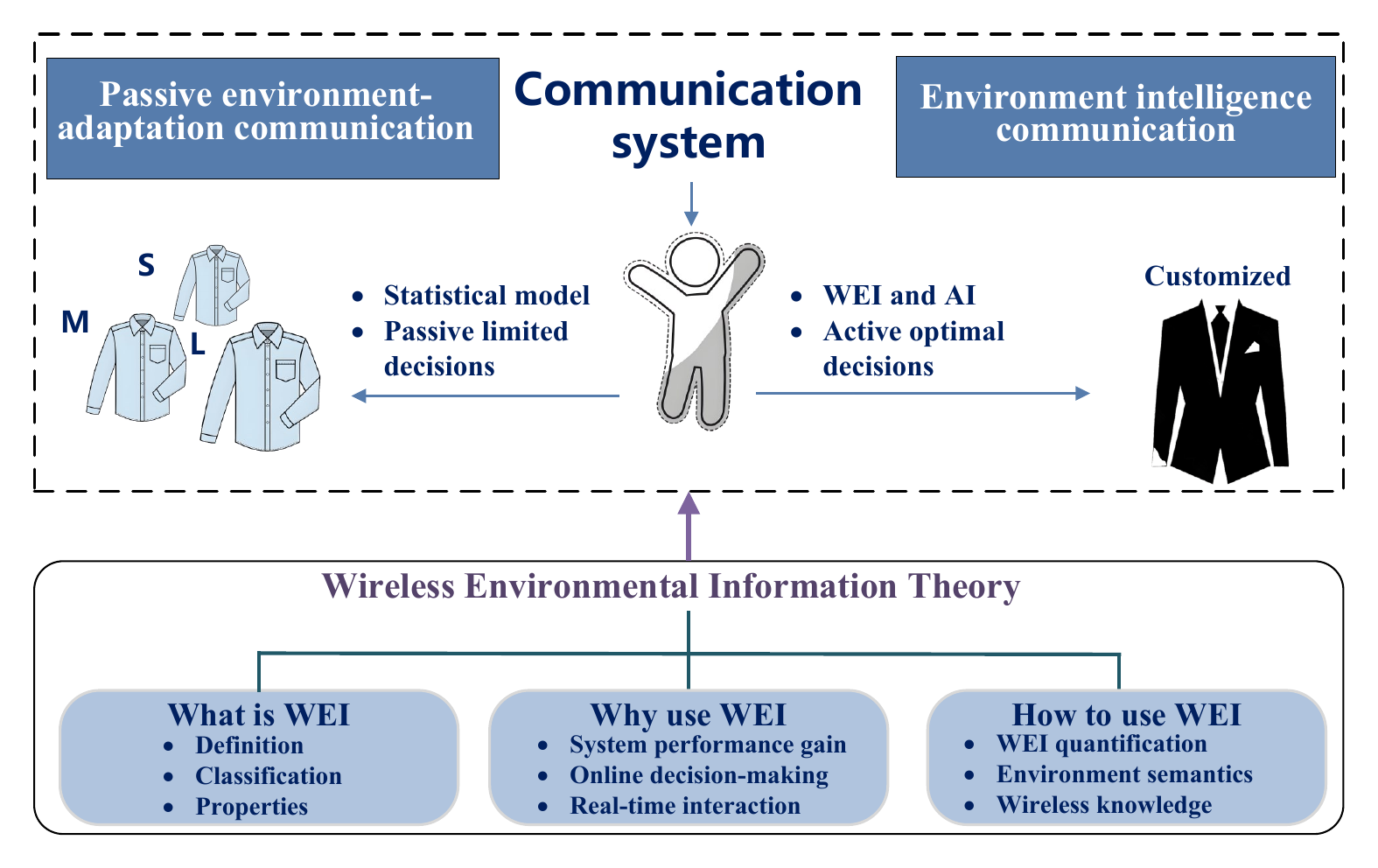}}
	\caption{WEIT: From passive environment-adaptation to environment intelligence communication.}
	\label{Environment Intelligence}
\end{figure*}
\subsection{Wireless environmental information}

Fig. \ref{Environment Intelligence} vividly depicts the similarity between channel acquisition for a communication environment and a customer purchasing apparel in a shop. As different body shapes necessitate corresponding clothing sizes, various environments require distinct channel models. Comparable to garment sizes including large, middle, small sizes; the standardized communication environments are finite, such as UMi, UMa, and RMa \cite{3Gpp38901}. This classification mode facilitates rapid access to clothes fit from the perspective of statistical sizes; however, people may not find a clothes size specially designed for each individual, especially someone with extreme tall or small figures. Environment intelligence communication, resembles high-end customized tailoring, where detailed measurements of an individual’s body are taken with measuring tapes, to create the most fitting clothing. The base station’s multimodal sensing capabilities serve as the “measuring tape” for the communication system, capturing detailed geometry and distribution information for a specific environment to accurately obtain environmental information, while AI acts as the “tailor” to make the optimal decision to improve system performance. 
To develop and achieve the environment intelligence, it is a prerequisite to establish the fundamental theory for wireless environmental information.

\textbf{Definition of wireless environmental information: }
The properties of wireless environment determine the characteristics of wireless channel. Electromagnetic waves interact with the objects and scatterers in the physical environment, such as buildings, metallic vehicles and rough roads, plants and so on. Maxwell’s equations, along with the boundary conditions, characterize the spatial strength distribution and variation of the electromagnetic wave via electromagnetic parameters of the environmental objects, including permittivity, permeability, and conductivity. Therefore, the physical properties of objects and scatterers in the propagation environment which affect wireless channel characteristics are defined as WEI. The diverse application scenarios of 6G, ranging from the ocean to space satellites, render the scope of WEI correspondingly broad, including variables such as water fluctuations, alterations in air humidity, and satellite trajectories. 

\begin{figure*}[!t]
	\centerline{\includegraphics[width= 17.5 cm]{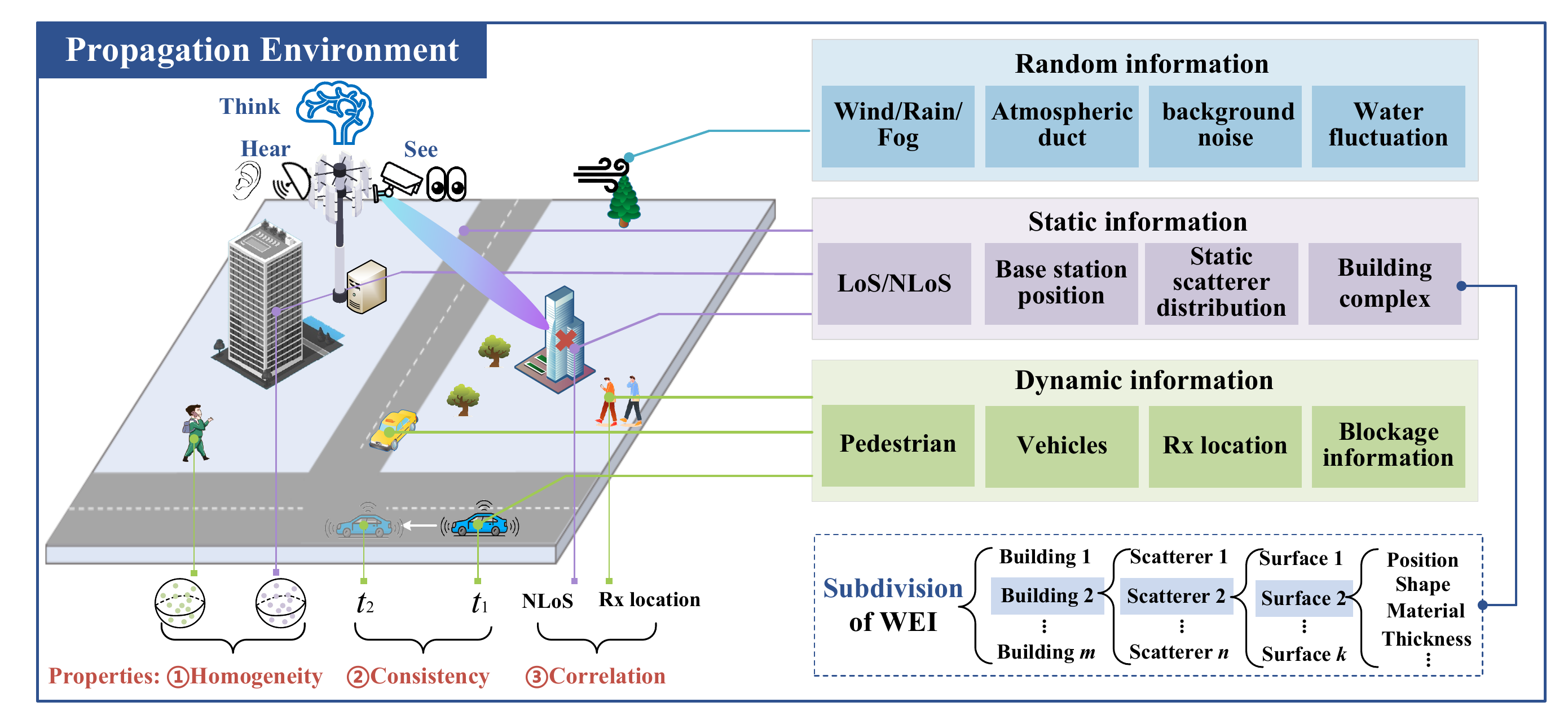}}
	\caption{Illustration for the WEI: definition, classification and properties.}
	\label{WEI}
	\vspace{- 2mm}
\end{figure*}

\textbf{Classification of wireless environmental information: } As shown in Fig. \ref{WEI}, WEI can be divided into three categories: static, dynamic, and random information. Static information keep stationary with fixed shapes during the observation time scale, including buildings, vegetation, streets, together with their height, outline, distribution, etc. Dynamic information exhibits temporal continuity and can be sensed and calculated, e.g., vehicles, pedestrian, scatterer mobility, and other mobile information with a distinguishable time scale. Random information such as the birth and death of scatterers caused by wind or tree shaking, and the background noise generated by electromagnetic interference and molecular thermal motion in the environment, which exhibit strong unpredictability in the time domain.

\textbf{Properties of wireless environmental information:}
It is worth noting that WEI possesses properties such as homogeneity, consistency, and correlation. Homogeneity refers to the fact that three types of WEI obtained by various sensing devices has the same attributes. For example, from the perspective of electromagnetic wave propagation, both static buildings and dynamic pedestrians can be modeled as scatterers, characterized by identical parameters including volume, shape, and material. Consistency refers to the stability and continuity of WEI across spatial and temporal dimensions. For instance, when an object continuously moves within the observation time $t_1$ to $t_2$ in the time domain, the corresponding WEI (such as position, velocity, and direction of motion) of this object can be considered consistent. Correlation indicates the interdependence between multiple WEI. For example, blockage information depends on the positions of the base station, user, and scatterers, and it changes as the user and dynamic scatterers move.

\subsection{Wireless environmental entropy definition}

From the preceding discussion, WEI will help to reduce channel variation uncertainty. Naturally, this prompts the inquiry: does an entropy associated with a WEI exist? For communication systems, the environment can be assumed to be random. The wireless environmental entropy serves to quantify the level of uncertainty in a stochastic wireless environment. In a given communication environment, the environmental entropy should have an upper bound ${{S}_{e}}\left[ \max \right]$, and a lower bound ${{S}_{e}}\left[ \min \right]$. For the upper bound, the possible objects within a certain area (communication system service range) are finite, and the number of possible environments composed of these objects is also finite, thus the environmental entropy should have a maximum value ${{S}_{e}}\left[ \max \right]$; The lower bound ${{S}_{e}}\left[ \min \right]$ mostly arises from environmental measurement abilities and errors, resulting in uncertainty within the environment, similar to the implications of the Cramér-Rao bound \cite{CRLB1,CRLB2}. 
Any WEI can be characterized by its dimension $d$ and quantity $\vartheta$, expressed as $\xi(\vartheta) = d \cdot \vartheta$. From a dimension's perspective, spatial position can be quantified as a three-dimensional vector. From the quantity's perspective, the precision of the position information is related to its measurement accuracy. A high-precision WEI ${{\xi }_{2}}(\vartheta)$ contains more information than a low-precision WEI ${{\xi }_{1}}(\vartheta)$. Thus, as measurement mean squared error (MSE) decreases, the amount of WEI increases. Fig. \ref{WEI} further demonstrates the detailed subdivision method for WEI. 
Taking a building complex as an example, the complex contains $M$ buildings, among which the $m$-th building contains $N$ scatterers, and the $n$-th scatterer is composed of $K$ surfaces. The information related to electromagnetic wave propagation on each surface can be divided into $i$ categories, such as position, shape, material, thickness, etc. The total amount of WEI that this building complex can provide is $M\times N\times K\times \sum\limits_{i}{\xi_{i} (\theta )}$.

\begin{figure*}[!h]
	\centerline{\includegraphics[width = 6.25in]{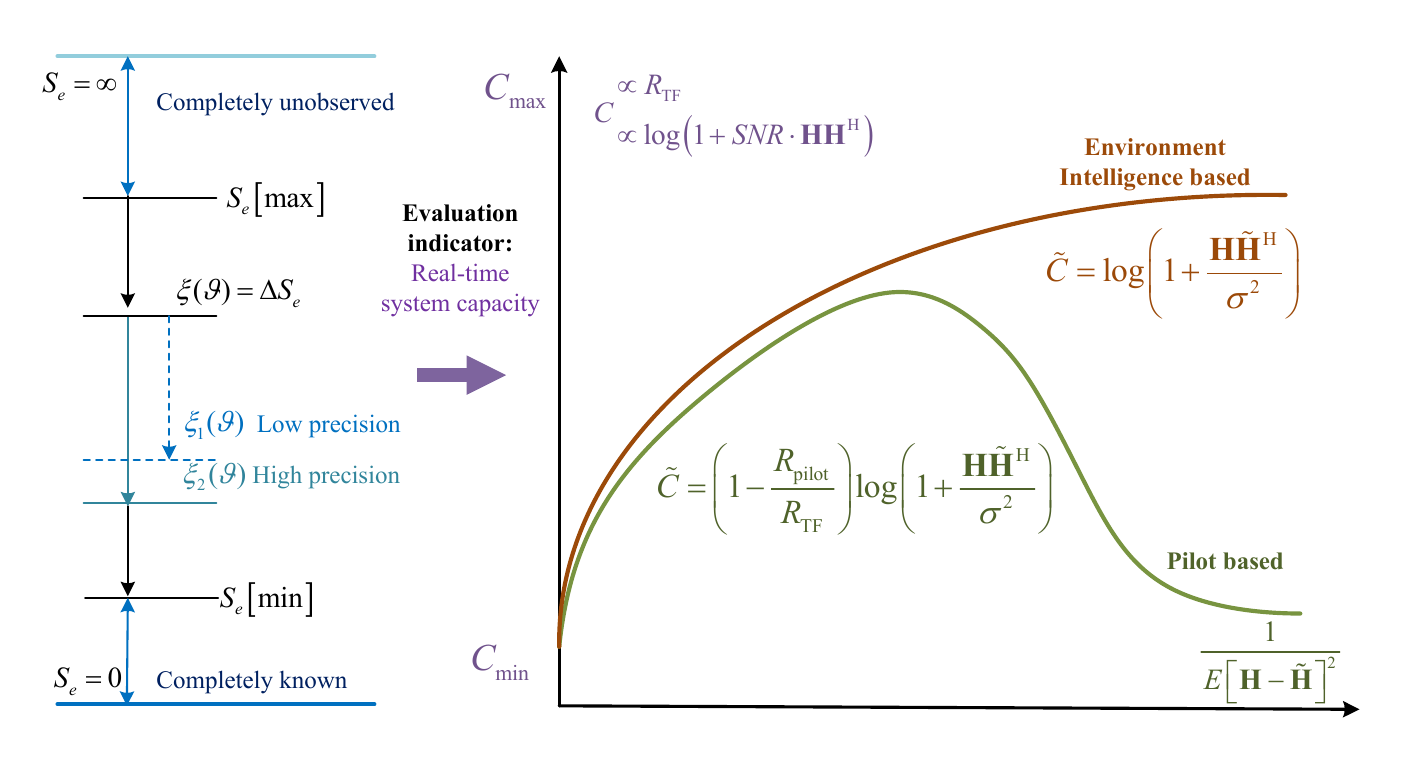}}
	\caption{The relationship between WEI and environmental entropy.}
	\label{entropy}
\end{figure*}

Evaluating environmental entropy from the perspective of system capacity is crucial for communication systems. Fig. \ref{entropy} also demonstrates the sources of system capacity improvements realized through environment intelligence communication compared to traditional communication. The system capacity is directly related to the quality of the channel and the allocated time-frequency resources. With the same time-frequency allocation, better channel quality results in higher system capacity; Similarly, with fixed channel quality, allocating more time-frequency resources leads to greater capacity. Traditional communication systems that rely on statistical models which are based on the statistical averages of selected typical environments, inadequately approaches the instantaneous optimal upper limit of performance. Traditional systems necessitate a trade-off to obtain more accurate channel information, utilizing a portion of time-frequency resources, such as pilot signals, to acquire real-time channel state information. This establishes a balance between time-frequency resources and channel quality, thereby constraining the system capacity. In contrast, environment intelligence communication leverages sensing and AI to acquire accurate channel information. The ``sensing resources'' operate independently of the time-frequency resources allocated for communication, thereby enabling the system to maximize its utilization of these essential time-frequency resources.

As previously stated, in communication systems, the relevant information of interest is related to electromagnetic wave propagation. The precision and property of information are both critical considerations. For example, compared to the mass of objects, the system is more concerned with the dielectric constant. Therefore, it is necessary to restrict the maximum types of WEI that a system can obtain, while also addressing accuracy issues. WEI serves as the bridge between the propagation environment and the channel, and a decrease in environmental entropy means an increase in the channel determinacy. A mapping relationship $\mathcal{F}\left( \cdot  \right)$ to process the obtained WEI into various channel parameters. Different WEI is transformed into specific channel parameters, such as delay, Doppler shift, power, via the mapping relationship $\mathcal{F}\left( {{\xi }_{1}},{{\xi }_{2}},\ldots ,{{\xi }_{i}} \right)$. 

\section{Environment intelligence communication aided by WEI}\label{secIII}
To overcome the limitations of the statistical paradigm and enable a communication system that interacts in real-time and utilizes WEI online, in this section, we propose a WEI based environmental intelligent communication framework. The motivation for introducing intelligence into the communication system was elaborated in detail, and the use of WEI was graded.
\subsection{Three cornerstones for EIC}

Compared with 5G, 6G wireless channels will continue to expand in terms of frequency, application scenarios, and support technologies. This requires precise capture of new features and characteristics of the channel, and low complexity integration into the theoretical framework of 6G. Channel prediction based on the online acquired WEI is a relatively low-cost method for channel acquisition. Therefore, the following three techniques will play an important role in extracting WEI and predicting channel changes.

\textbf{Multimodal environment sensing:} ISAC technology can enable smarter and more efficient environmental sensing and WEI acquisition. Meanwhile, base stations are progressively incorporating various sensing devices such as industrial sensors, depth cameras, LiDARs, and millimeter-wave radar. It can collect multimodal sensing data from different angles, locations, and scales, thereby providing a comprehensive perception of the environment. For example, millimeter-wave radar can provide precise information on the position and velocity of objects, LiDAR can offer high-precision spatial data, and color cameras can capture material information about objects. These sensing data can be efficiently fused, enabling accurate capture of full-dimensional WEI for both moving objects and static scatterers, thus providing a data foundation for channel fading prediction.

\textbf{AI enabled online channel prediction:} Traditional channel fading derived from empirical and statistical models typically relies on a limited set of typical environments, making it insufficient to fully capture the complexity of wireless channel variations.
With the rise of AI technologies, particularly advancements in machine learning and deep learning, AI has become a crucial tool for tackling this issue. By fully utilizing WEI that encompasses comprehensive environmental states, AI models are better equipped to understand and predict the dynamic changes in wireless channels. With their powerful feature extraction and nonlinear mapping capabilities, AI models can accurately uncover knowledge from vast amounts of WEI, learning and identifying potential patterns in channel fading, thereby providing valuable insights for pilot cost reduction, beam management, resource optimization and so on.

\textbf{Wireless environment knowledge construction:} With the improvement in measurement device resolution and the increasing number of sensing devices, wireless channel data will inevitably exhibit significant big data trends across various environments. A large volume of channel data forms the foundation for uncovering hidden patterns, however, this also increases the computational complexity of channel fading prediction, limiting the real-time performance of inference. To reduce data volume and extract channel-relevant information, we propose WEK as a representation of the mapping relationship between the environment and the channel \cite{REKP}. This abstracts the channel propagation process from both the environment and the channel, enabling intelligent understanding of the underlying propagation laws through theoretical derivations, mathematical formulations, or semantic interpretations.

\newcommand{\tabincell}[2]{\begin{tabular}{@{}#1@{}}#2\end{tabular}}
\begin{table*}[h]
	\centering
	
	\caption{The characteristics of different channel acquisition methodologies from system perspective.}
	\begin{tabular}{c|c|c|c|c|c|c}
		
		\hline
		& \textbf{Method} & \textbf{Real-time} & \textbf{\makecell{WEI \\utilization}} & \textbf{\makecell{WEI \\ acquisition}} & \textbf{Complexity/Cost} & \textbf{Tasks} \\ \hline 
		\multirow{2}{*}{\textbf{\makecell{Channel \\ modeling}}} & Statistical model & \multirow{2}{*}{Offline} & Level 2 & \makecell{Channel\\ measurement} & Low & \multirow{2}{*}{System deployment} \\ \cline{2-2} \cline{4-6}%
		& Deterministic model &   & Level 1 & \makecell{Environmental\\ measurement} & High &  \\ \hline
		\textbf{\makecell{Channel \\ estimation}} & Pilot/Pre-training & Online & Level 2 & Pilot & High & \makecell{Obtain current \\ channel parameters} \\ \hline
		\multirow{2}{*}{\textbf{\makecell{Channel \\ prediction}}} & Neural network & \multirow{2}{*}{Online} & Level 3 & Historical CSI & Acceptable & \multirow{2}{*}{\makecell{Predicting\\ channel parameters}} \\ \cline{2-2} \cline{4-6}
		& Signal processing &  & N/A & Current signal & High &  \\ \hline
		\textbf{\makecell{Environment\\ intelligence}} & WEK+AI & Online & Level 4 & Sensing & \makecell{ High (building) \\ Low (usage)}
		& \makecell{Task oriented\\ intelligent interaction}  \\ \hline
		
	\end{tabular}
	\label{table_model2pre}
\end{table*}

\subsection{Different levels to acquire CSI from system perspective}
The acquisition and processing of WEI can be categorized into 4 levels, progressively reflecting the evolution from manual operation to fully automated intelligent sensing:

\textbf{Level 1:} At this stage, the acquisition of WEI relies entirely on manual operations. Environmental measurements are conducted manually, and the collected data is used to reconstruct a complete 3D environment model by hand. This offline WEI acquisition approach is inefficient and incapable of meeting the demands of large-scale dynamic environments.

\textbf{Level 2:} While the collection of environmental data still depends on manual operation, the data processing methods have been optimized. After data collection, key features are extracted either manually or through automated techniques such as AI algorithms, enabling low-dimensional feature extraction and data organization. For example, complex 3D environmental model is simplified into a parameter set consisting of specific features such as the height of base station, scene size, and the number of buildings. 

\textbf{Level 3:} At this level, the data collection process becomes automated. Various sensing devices deployed in the environment automatically collect WEI. AI algorithms are then employed to automatically extract the environmental features necessary for channel prediction. This stage eliminates the limitations of manual data collection, significantly improving the efficiency of data acquisition and enabling adaptation to more complex environmental changes. However, the reliance on neural network-based feature extraction introduces a limitation in the interpretability of WEI.

\textbf{Level 4:} On the foundation of fully automated multimodal sensing and data fusion, the environment-specific features and channel characteristic parameters are explicitly related using WEK. This enables interpretable extraction of core features required for subsequent tasks. This stage not only achieves real-time WEI acquisition but also allows for concise and accurate representation of channel characteristics, combining efficiency with interpretability.
\begin{figure*}[!h]
	\centerline{\includegraphics[width= 17cm]{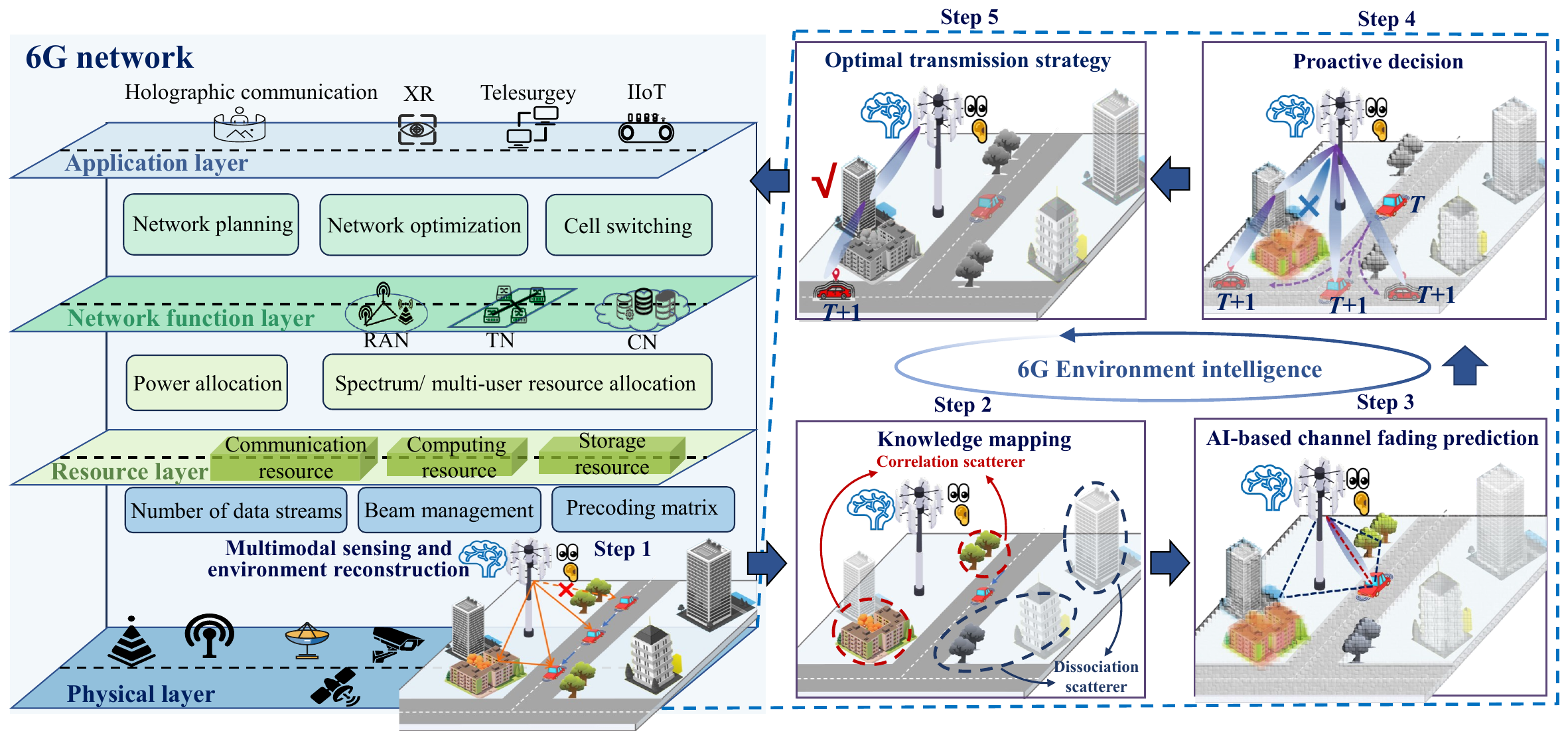}}
	\caption{Environment intelligence communication for 6G system.}
	\label{EIC}
	\vspace{-3mm}
\end{figure*}

Since the advent of wireless communications, extensive research is being conducted on propagation environments. WEI is now making its presence felt in various domains, including channel modeling, channel estimation, and channel prediction. The use of WEI is gradually transitioning from offline to online and from vague to precise. In both deterministic and statistical models, the utilization of WEI in channel modeling remains relatively basic. Channel estimation schemes currently rely on pilots to capture channel variations, requiring time-frequency resource overhead, which encroaches on communication resources. Early applications of AI for channel prediction primarily use environmental information offline, employing statistical learning on historical channel data to predict future channel states. In comparison, traditional signal processing focuses more on processing past instantaneous information rather than forecasting future channel variations. With the theoretical support of WEIT, environment intelligence now enables the real-time, online utilization of WEI, ensuring communication resources are preserved while achieving accurate channel fading predictions.
Table \ref{table_model2pre} provides a conceptual comparison of WEI between different channel concepts.

\subsection{Architecture for the proposed EIC-WEI }
Environment intelligence communication is a closed-loop process. It begins with leveraging sensing methods to obtain accurate and comprehensive WEI based on the task requirements of the communication system. Next, use WEI to predict the various channel fading, and finally, the communication system autonomously determines the optimal transmission strategy, enabling intelligent interaction.

\textbf{Step 1 multimodal sensing and environment reconstruction:}  By intelligently integrating multimodal sensing data, the system automatically analyzes information collected from various sensing devices to reconstruct a high precision 3D model of the entire scene. In this process, devices such as cameras and LiDAR act like human eyes, accurately capturing the structure of the surrounding environment, while radar acts as human ears, acquiring the velocity and position of scatterers by "listening" to the echo signals. This enables the reconstructed environment model contains all critical details of the propagation environment, providing comprehensive environmental information for the communication system.

\textbf{Step 2 knowledge mapping:} For a specific communication task, the complete 3D model often contains redundant and complex information. Therefore, it needs to be processed based on the task requirements to extract usable WEI. These requirements can originate from various layers of the network, including the physical layer, resource layer, and others. The WEI extraction methods involve noise reduction, removal of irrelevant data, and ultimately mapping the information into a form suitable for the prediction layer. More details will be elaborated in the next subsection.

\textbf{Step 3 AI-based channel fading prediction:} Leveraging trained neural networks, the processed WEI is utilized to selectively predict channel parameters based on the communication task requirement, thereby reducing overhead. At this stage, WEI is employed in an online manner, and the final prediction results will directly influence system decisions.

\textbf{Step 4 proactive decision:} User actions often follow multiple potential paths, requiring the communication system to make proactive decisions tailored to each environment. As shown in Fig. \ref{EIC}, a vehicle approaching an intersection may choose to turn left, go straight, or turn right. Here, like the human brain, AI-enabled EIC can make fast and optimal strategy decision based on multiple possible channel predictions of user moving status, thereby the system performance is improved intelligently.

\textbf{Step 5 optimal transmission strategy:} Once the user’s actual choice is made, the communication system applies the corresponding optimal strategy from the previous step. The process then proceeds to the next round of task requirements.
\subsection{The WEI flow: unlocking the EIC}

\begin{figure*}[h]
	\centerline{\includegraphics[width = 17.4cm]{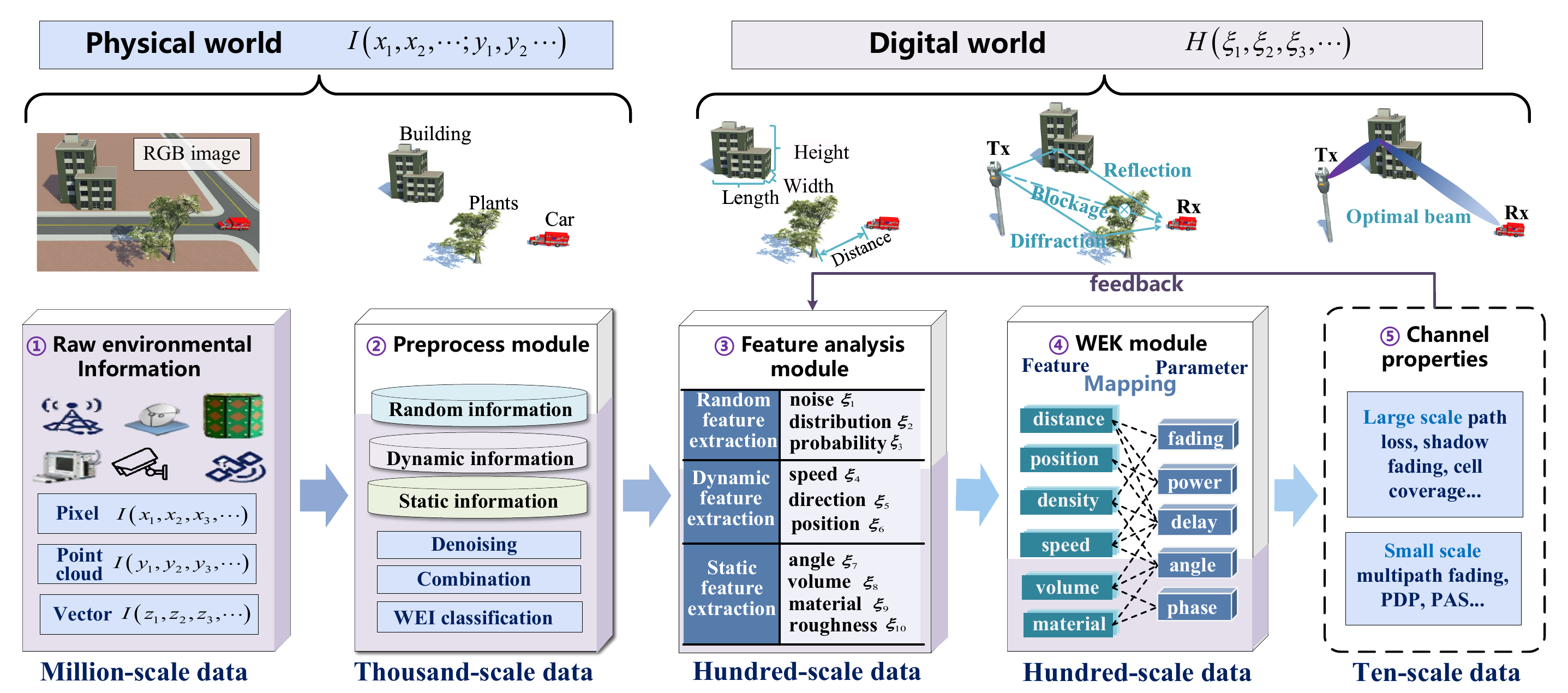}}
	\caption{WEI five steps flow for EIC: from raw data to WEK.}
	\label{Framework}
	\vspace{-3mm}
\end{figure*}

In the closed-loop process of environment intelligence, it is very important to recognize and utilize WEI as it is the bridge between the environment and the channel. Now, focusing on Fig. \ref{Framework}, we will provide a detailed explanation of each module in the WEI flow processing.

1) WEI collection module: From the perspective of WEI acquisition, we can divide it into two categories: visual sensitive information such as volume and electromagnetic sensitive information such as scattering coefficients, which cannot be fully captured by a single device. For instances, the utilizing of camera is a conventional means for collecting WEI. Exploiting computer vision technologies, such as object detection, object tracking, and semantic segmentation, one can identify and extract environmental objects from images; And radar can determine the position of objects through echoes. Through the using of multi-modal sensing devices, we can acquire a vast amount of raw wireless environmental information in various forms, typically measured in the millions.

2) WEI preprocessing module: The raw wireless environmental information contains a massive amount of multimodal data along with significant ambient noise, necessitating processes like denoising, integration, and classification. The processed wireless environmental information will be divided into three categories: static WEI, dynamic WEI  and random WEI. Each category of WEI is then processed using distinct workflows tailored to its characteristics.

3) Environmental feature analysis module: The processed WEI will undergo feature analysis to extract the parts of the channel that are affected by different information. Machine learning and data mining technology can be used to achieve efficient feature extraction and reduce data dimension. In this process, three types of WEI are processed separately, but the extracted features may be similar. For example, buildings can be simplified to position, size, and material, while dynamic vehicles can extract features such as position, material, and speed. This module significantly reduces the data dimension while providing structured input for subsequent processes.

4) Wireless environment knowledge module: To quickly and accurately obtain channel characteristics, wireless environment knowledge should be established in advance to abstract the channel propagation process from environment and channel. This enables AI algorithms to understand and learn the underlying mapping rules of channel behavior, facilitating accurate predictions of channel fading states. These predictions can be utilized to optimize transmission techniques, significantly enhancing system performance \cite{REKP}.

5) Channel characteristic module: The processed channel parameters will be mapped to different channel characteristics at this module, including large-scale and small-scale characteristics. Different channel characteristics can be mapped separately to improve system efficiency. The channel characteristic module is closely connected to the preceding WEI processing stage. To ensure system efficiency, feedback information should be provided to the environmental feature analysis module, enabling targeted predictions and computations. The feedback information should also be translated into simple control commands to meet real-time requirements.

\section{Task-oriented EIC-WEI validation from theory to practice} \label{secIV}

\begin{figure*}[!h]
	\centerline{\includegraphics[width = 15.5 cm]{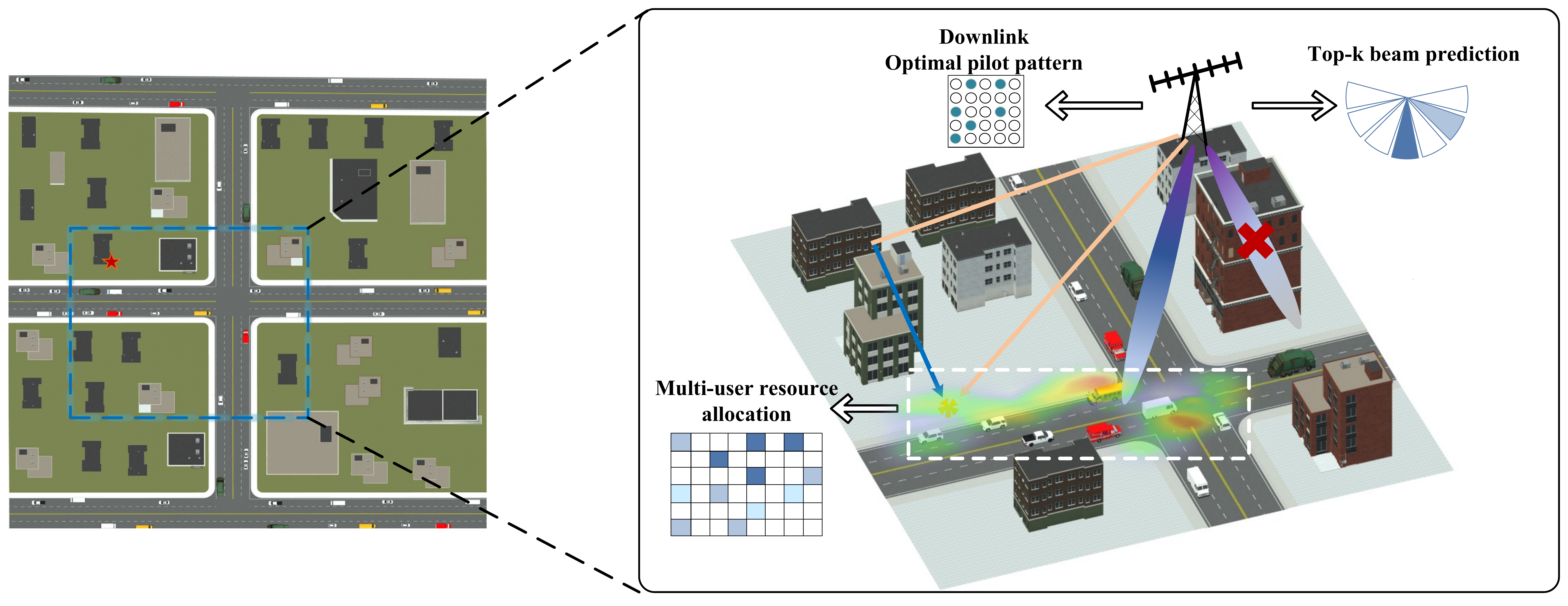}}
	\caption{ 2D perspective of simulation test environment A and specific tasks.}
	\label{scenario}
\end{figure*}
\begin{figure*}[!t]
	\centerline{\includegraphics[width = 17 cm ]{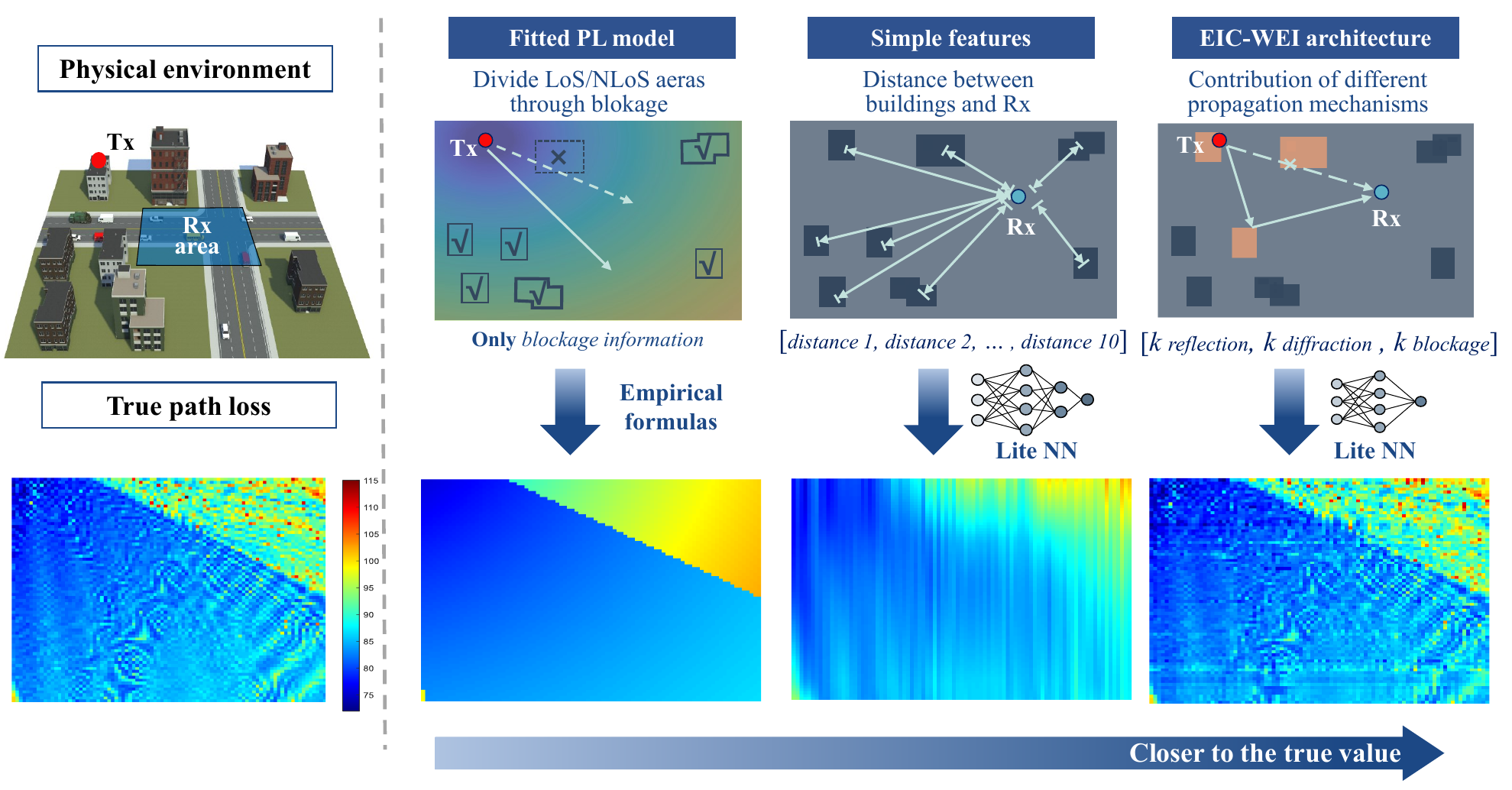}}
	\caption{Cell coverage results based on the true channel statistical channel model, simple features, and EIC-WEI.}
	\label{GBSM_pre}
	\vspace{- 1mm}
\end{figure*}

In this section, to evaluate the performance of environment-enhanced channel prediction, we will discuss the gains that the EIC-WEI architecture brings to the communication system through four tasks-cell coverage, CSI prediction, beam selection—within the same test environment, and resource management in a new test environment to validate the generalization. The simulation test environment is shown in Fig.~\ref{scenario}, where the location of the transmitter (Tx) is marked with an asterisk at a height of 19 meters, and the single-antenna users are evenly distributed across four lanes. The constructed outdoor urban environment is built by the autonomous driving simulation software CARLA \cite{CARLA}, with four building groups, and different types of vehicles are randomly placed on one side of the road to ensure that the environment has enough diversity. As the loss of surface detail has a limited effect on the channel, the entire test environment is imported into blender software for simplification, where buildings and vehicles are replaced with simple cubes. The simplified model of the environment is imported into WirelessInsite for ray tracing simulation. The test environment is a square with a side length of 200 meters. Detailed parameter settings can be found in Table \ref{parameter}. Wireless environmental information includes panoramic image (segmented into four directions) of the receiver (Rx) and the coordinates of buildings within the environment. All datasets are generated by the Beijing University of Posts and Telecommunications and China Mobile Communications Group-DataAI-6G Dataset (BUPTCMCC-DataAI-6G Dataset) \cite{database1, database2}. 

\begin{table}
	\centering
	\caption{Simulation parameter settings}
	
	\begin{tabular}{>{\centering\arraybackslash}m{3.5cm}|>{\centering\arraybackslash}m{3.5cm}}
		\hline
		Parameters & Setting \\
		\hline
		Modulation & OFDM \\
		Middle frequency & 6775 MHz \\
		Bandwidth & 8.28 MHz \\
		Subcarrier & 120 kHz \\
		OFDM symbol number & 3 \\
		Tx antenna number & 128 \\
		Rx antenna number & 1 \\
		Tx location & [-57.4 m, 27m, 19 m] \\
		Rx set & height: 2 m gap: 0.25 m \\
		Panoramic image & 600 * 200 pixels  \\
		\hline
	\end{tabular}
	
	\label{parameter}
\end{table}

\begin{figure}[!h]
	\centerline{\includegraphics[width= 8cm]{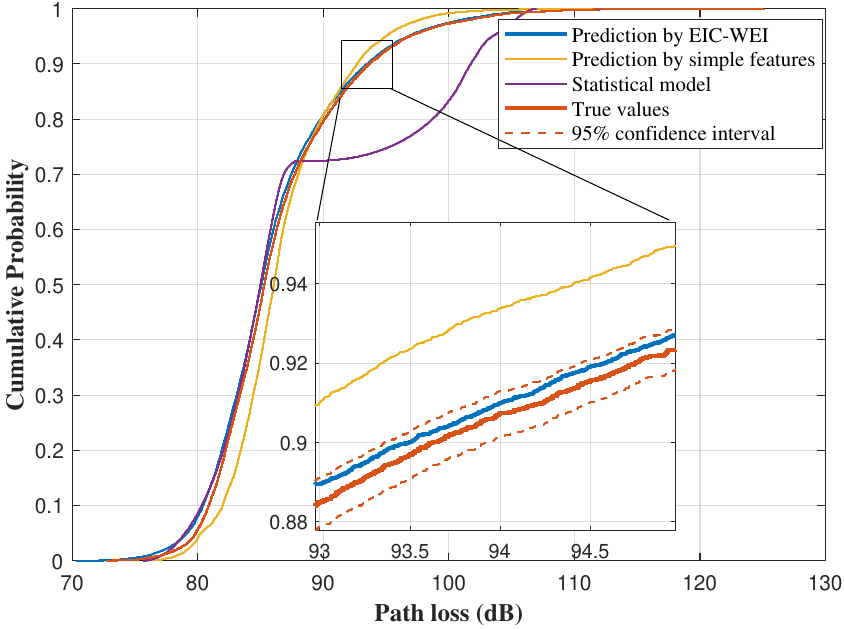}}
	\caption{CDF plots of the proposed WEI-based prediction and the contrast methods.}
	\label{CDF}
\end{figure}

\textbf{Task 1: cell coverage} 

The Rx area verifying cell coverage can be found in Fig. \ref{scenario}. It is evident that within the selected area, there is a tall building to the right of the Tx, which creates a significant blockage. Fig. \ref{GBSM_pre} shows the results of different channel acquisition methods for predicting cell coverage of channels at 6.775 GHz. The statistical channel model uses different empirical formulas in the line-of-sight (LoS) and non line-of-sight (NLoS) regions, and it can be seen that the prediction results have obvious regional divisions.
However, the simplistic relationship between channel fading and propagation distance results in severe distortion  \cite{3Gpp38901}. 
Channel prediction based on simple environmental features has significant shortcomings in determining direct areas. Compared to the previous two methods, the results of EIC-WEI are closer to the true values, demonstrating good fitting. The introduction of multimodal environmental information allows AI to determine which environmental features will affect electromagnetic propagation characteristics, thus achieving better results than predictions based solely on simple environmental features.

\textbf{Task 2: large/small scale parameter channel prediction}

Fig. \ref{CDF} demonstrates the cumulative distribution functions (CDF) of the above method for predicting the path loss of the channel. The prediction results of the statistical channel model show obvious segmentation due to the use of different empirical formulas for LoS and NLoS region; Channel prediction based on simple features more closely aligns with the trend of the true value compared to the statistical channel model. However, there remains a noticeable deviation from the true value. Compared to the first two methods, the results of EIC-WEI are closer to the true values, with the vast majority of results within the 95\% confidence interval.

\begin{figure}[!t]
	\centerline{\includegraphics[width= 8.5cm]{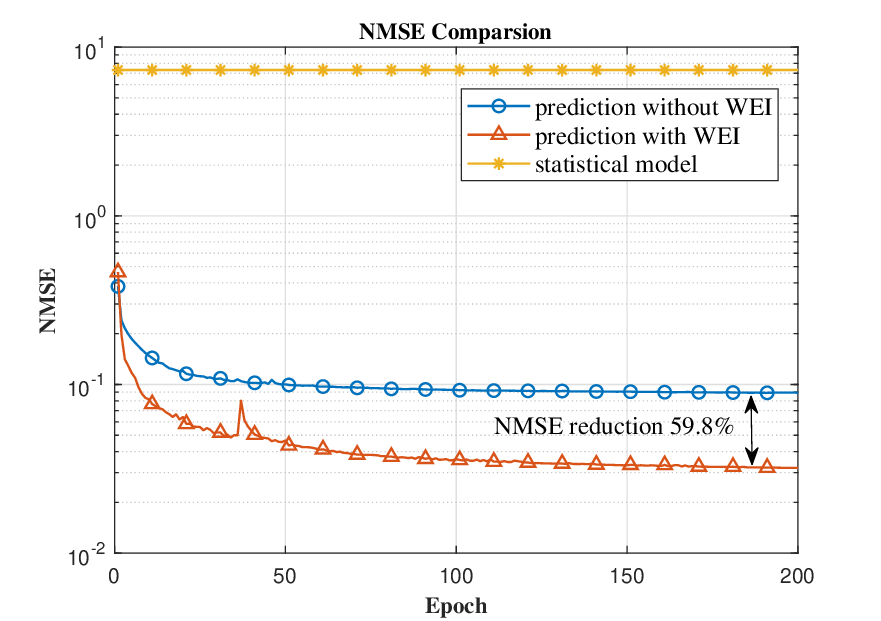}}
	\caption{The NMSE of the proposed WEI-based prediction, and GBSM methods of the channel.}
	\label{NMSE}
\end{figure}

Fig. \ref{NMSE} illustrates the performance of the proposed method in capturing small-scale channel characteristics. We use the normalized mean square error (NMSE) as an evaluation indicator and employ 1/8 of the resources for pilot signals. The channel modeling method is based on TR 38.901 \cite{3Gpp38901}, although this method incurs no additional overhead, its NMSE shows a significant gap compared to several other methods, that the channel model itself is not suitable for real-time communication tasks. The EIC-WEI architecture uses a panoramic image around the receiver as input for WEI. Compared to the prediction results without WEI, after the convergence of the network model training, the NMSE decreased by approximately 59.8\%, highlighting the potential of this architecture for predicting small-scale parameters.
\begin{figure}[!t]
	\centerline{\includegraphics[width= 8.5 cm]{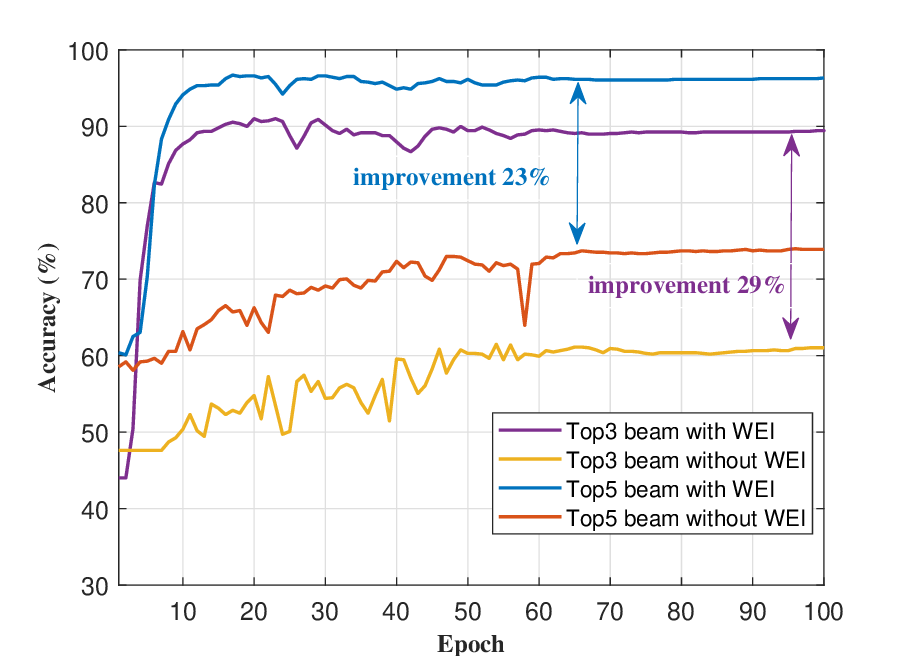}}
	\caption{Optimal beam prediction based on WEI.}
	\label{beam}
\end{figure}

\textbf{Task 3: optimal beam selection}

In addition to predicting channel parameters, WEI can also optimize channel beam prediction. In this task, beam codebook selection is performed at the Tx, which is equipped with 32 codebook options, aiming to maximize the received power. The prediction without WEI relies on past CSI, whereas the prediction with WEI utilizes panoramic images and CSI. Significant results can be observed from Fig. \ref{beam}: the prediction accuracy of the top 5 beams with the highest power improves by 23\%, while the accuracy of the top 3 beams increases by 29\%. Additionally, the inclusion of WEI leads to faster and more stable convergence. The result demonstrates the potential of the EIC-WEI.

\textbf{Task 4: optimization of air interface resources}

To optimize multi-user service in a typical vehicle-to-infrastructure (V2I) test environment, we simulate a resource allocation problem based on fairness. Ten terminal users are randomly selected as the served users on the main road, marked by the dashed box. Through the sensing capabilities of the base station, the system can autonomously track the movement of user locations. Given the significant differences in the communication environments across various user positions, our objective is to ensure the best possible service quality for each user. 
To achieve this, the objective of the resource fairness optimization is to maximize the minimum throughput of users, which can be formulated as $\max\limits_{\boldsymbol{X}_{u},u=1,2,...,N} T_{min}$. Here, $T_{min}=\min\limits_{u=1,2,...,N}\left\{T_u\right\}$ and $T_u=\sum_{t=1}^{T}{\sum_{r=1}^{R}{D_{u,t,r}}\cdot\boldsymbol{X}_u[t,r]}$. $D_{u,t,r}$ denotes The data rate transmitted by user $u$ in the $t$-th time domain and $r$-th frequency domain resource block. $\boldsymbol{X}_u \in \left\{0,1\right\}^{T\times R}$ denotes the resource allocation matrix for user $u$, with dimensions $T\times R$, where each element takes values from the set $\left\{0,1\right\}$. The constraint is satisfied for all $\boldsymbol{X}_u$ such that $\sum_{u=1}^{N}{\boldsymbol{X}_u}=\boldsymbol{1}_{T\times R}$, where
$\boldsymbol{1}_{T\times R}$ represents a matrix of dimensions $T\times R$ with all elements are equal to $1$.

\begin{figure}[!t]
	\centerline{\includegraphics[width = 2.7 in]{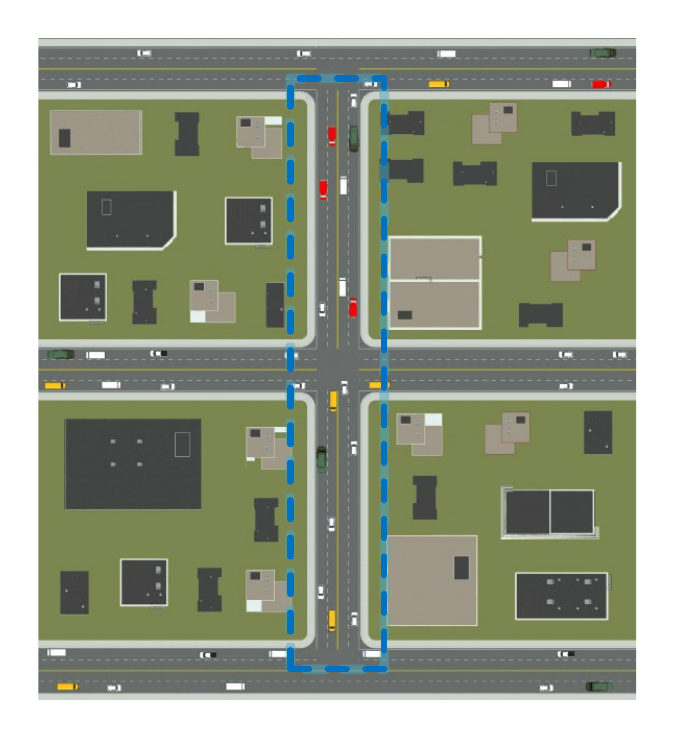}}
	\caption{Simulation test environment B: generalization verification.}
	\label{scenario2}
	\vspace{-5mm}
\end{figure}

\begin{figure}[!t]
	\centerline{\includegraphics[width = 8.5 cm]{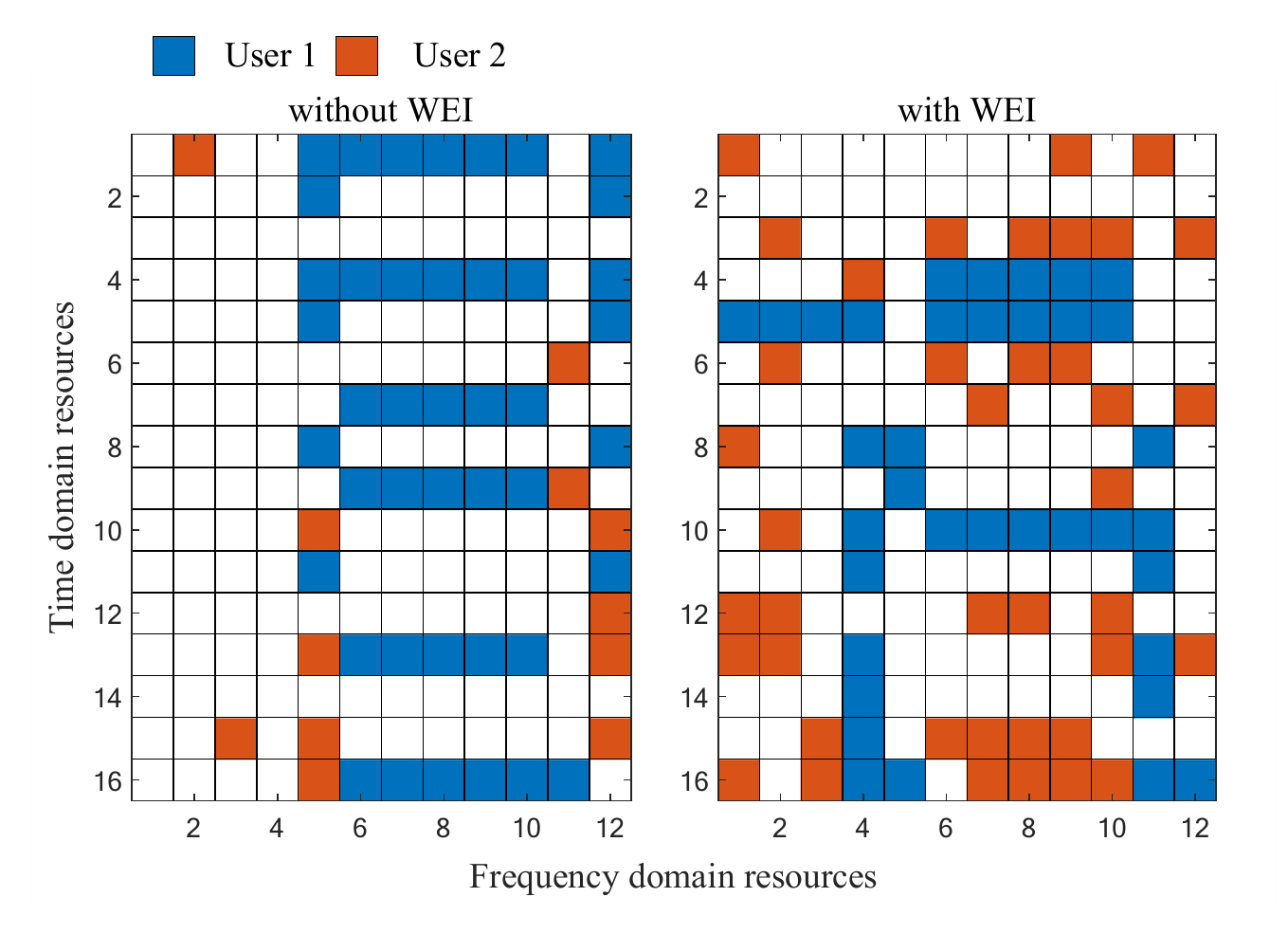}}
	\caption{Time-Frequency resource block allocation for two users based on WEI.}
	\label{resource1}
	\vspace{-5mm}
\end{figure}

\begin{figure}[!t]
	\centerline{\includegraphics[width = 8.5 cm]{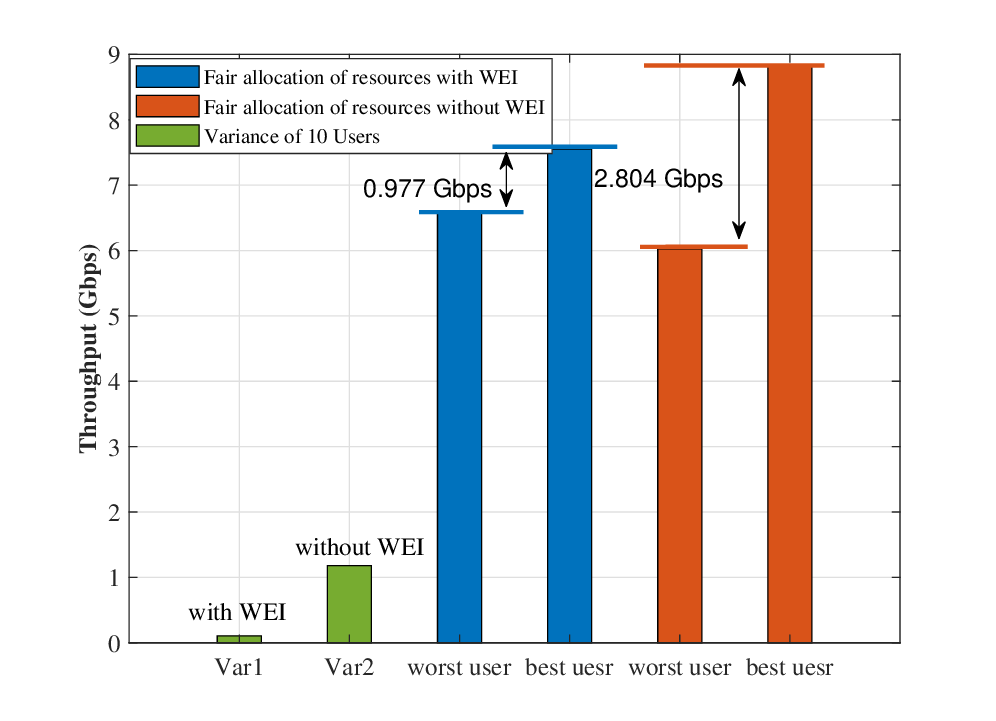}}
	\caption{Throughput variance of 10 users and throughput difference between the best and worst users.}
	\label{resource2}
	\vspace{-5mm}
\end{figure}

The goal of the this task is to allocate resources more balanced among the ten users within the blue dash line, with each antenna transmitting at a power of -8 dBm. To further verify the generalization of the proposed method, we use the CSI prediction model trained in environment A to test the performance in a new environment B, as shown in Fig. \ref{scenario2}. Before incorporating WEI for resource allocation, it can be observed that User 1 receives significantly more resources than User 2. After incorporating WEI, the time-frequency resource allocation between User 1 and User 2 becomes more balanced as shown in Fig. \ref{resource1}. Fig. \ref{resource2} presents the users' throughput. Before incorporating WEI, the throughput gap between the user with the lowest and the highest throughput is 2.804 Gbps. After resource allocation using WEI, this gap is reduced to 0.977 Gbps. The variance of throughput among the 10 users is 0.10 Gbps, and the total system throughput is 72.04 Gbps. In contrast, without WEI, the variance of throughput among the ten users is 1.18 Gbps, with a total system throughput of 72.21 Gbps. Fairness optimization among multiple users is achieved without sacrificing total throughput.

\section{Main challenges and future work}\label{secV}
\begin{figure*}[!h]
	\centerline{\includegraphics[width = 15 cm]{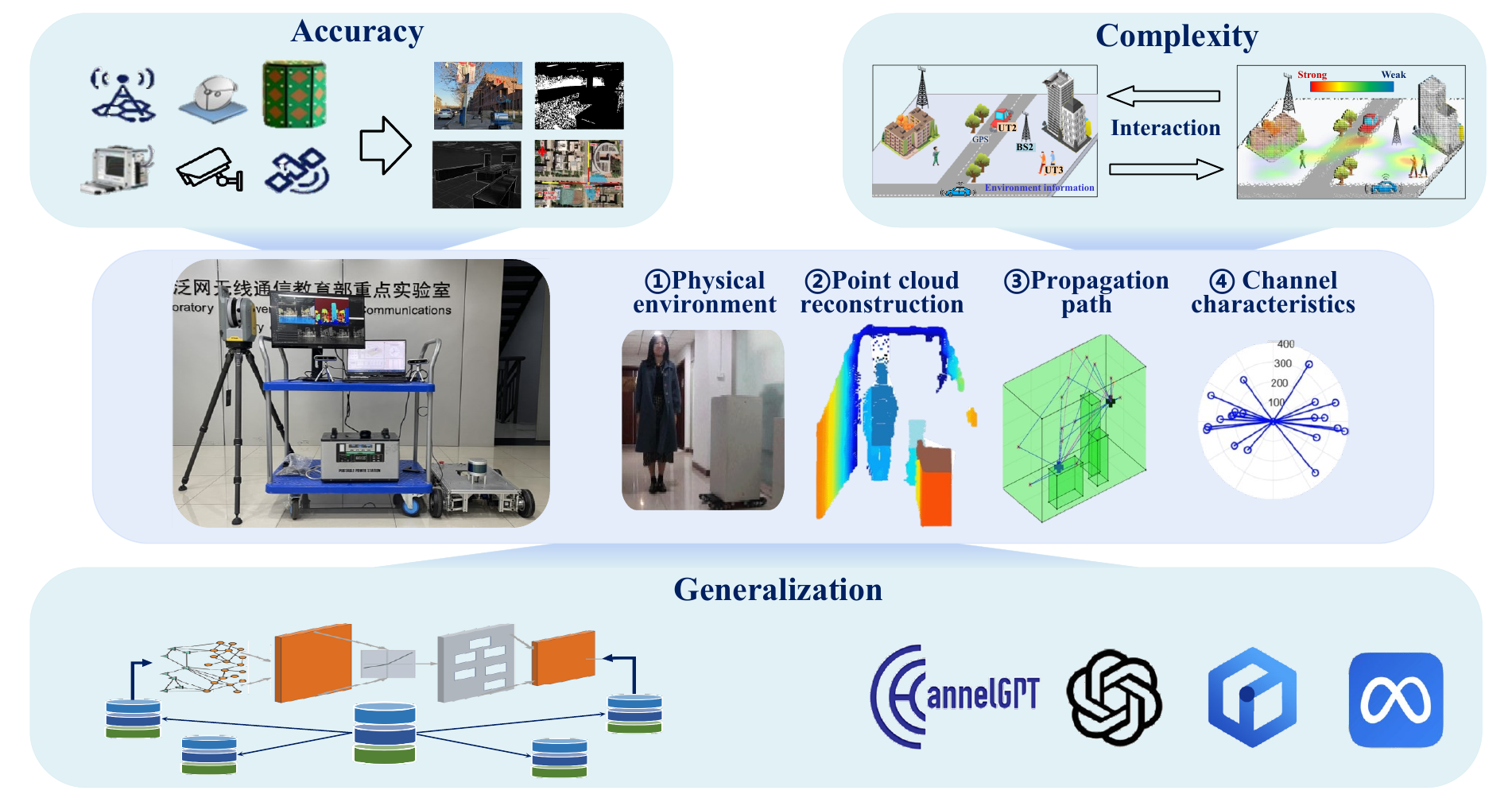}}
	\caption{Main challenges and prospects for future EIC system: accuracy, complexity and generalization.}
	\label{openissue}
	\vspace{- 4mm}
\end{figure*}
As shown in Fig. \ref{openissue}, we have developed a real-time interaction platform enabled by environment sensing as a prototype system for EIC. Specifically, this system leverages collected point cloud data to automatically extract scatterer features and perform multimodal fusion. Subsequently, it reconstructs environmental objects in real-time and predicts propagation paths, derives channel parameters based on propagation paths. This system demonstrates the feasibility of integrating sensing and AI for channel prediction. However, to achieve a fully intelligent and comprehensive EIC-WEI suitable for 6G, several challenges and open issues require further investigation.

\textbf{The accuracy of WEI acquisition to ensure system performance:}
High-fidelity environmental reconstruction remains a challenge in wireless environmental information collection. Accurate data collection is crucial for subsequent processing, yet current methods often struggle with precision. Moreover, sensing devices are showing a trend towards multimodal development, where multimodal sensing devices will bring multimodal WEI. Future digital twin platforms should support the joint processing of multimodal information. Besides, quantitatively describing the various types of collected WEI remains a challenging work.

\textbf{Low-complexity environment interaction to support real time tasks:}
Real-time interaction is a key requirement of EIC-WEI, with its core lying in the rapid sensing, processing, and response to the dynamic wireless environment. However, implementing real-time WEI acquisition, WEI processing, channel fading prediction, and decision-making within the base station is a complex and challenging task. The interaction between the system and the environment should be achieved with minimal resource overhead and the lowest possible complexity to meet the 6G network's demands for low latency and high efficiency. Additionally, real-time communication between different base stations should be considered to ensure optimal decision-making across the coverage area.

\textbf{The environmental generalization to achieve ubiquitous intelligence:}
Existing channel prediction networks face significant generalization challenges, requiring not only adaptability to diverse communication tasks but also robust performance across different environments. To address these challenges, the construction of WEK could offer a feasible solution. WEK has strong representational power, enabling the establishment of many-to-many mappings that link multiple environmental features with various channel parameters, thereby capturing the comprehensive impact of the environment on the channel. This mapping framework provides a unified description of channel characteristics, which can be adapted to different environments, thus overcoming the limitations of task- and environment-specific models. Moreover, channel large models demonstrate exceptional modeling capabilities, multitask learning potential, and superior generalization performance. These models adopt a multitask learning framework to integrate channel knowledge across different communication tasks, enhancing their adaptability to new tasks. Additionally, diversified training data augmentation allows the model to learn variations in channel characteristics across diverse environments, showcasing its adaptability. However, integrating the specific knowledge into large models remains an area that requires further exploration.

\section{Conclusions}\label{secVI}
Wireless channel plays an increasingly important role in communication systems. In this paper, we propose a novel paradigm of environment intelligence communication for 6G. To facilitate further research, we first introduce wireless environmental information theory, systematically defining, classifying, and discussing the properties of wireless environmental information and then derive the wireless environmental entropy from the perspective of system capacity. Next, we propose the EIC-WEI architecture, which consists of environment sensing and reconstruction, WEI processing and extraction, AI based channel prediction, and optimal transmission strategy decision, to implement online intelligent interaction with the physical environment. To further expose the function of WEI within the EIC, a detailed step by step processing flow from raw environment data to wireless knowledge is illustrated. Finally, successive numerical results are presented to verify the feasibility of our work, from several different-level tasks for a specific test environment. Some potential future research directions and promising application environments are discussed.

\section{Acknowledgment}
This work is supported by the National Key R\&D Program of China (Grant No. 2023YFB2904805), the National Natural Science Foundation of China (No. 62401084), and BUPT-CMCC Joint Innovation Center.

\section{Compliance with ethics guidelines}
Jianhua Zhang, Li Yu, Shaoyi Liu, Yichen Cai, Yuxiang Zhang, Hongbo Xing, and Tao Jiang declare that they have no conflict of interest or financial conflicts to disclose.

\end{document}